\documentclass[twocolumn,reprint,12pt]{iopart}
\usepackage{iopams}
\usepackage{graphicx}

\begin{document}
%\center
\title[]{Empirical tight binding parameters for GaAs and MgO with explicit basis through DFT mapping}%[Author guidelines for IOP journals in  \LaTeXe]

\author{Yaohua~Tan, Michael~Povolotskyi, Tillmann~Kubis, Yu~He, Zhengping~Jiang and Gerhard~Klimeck }
% \email{tyhua02@gmail.com}
%\affiliation{%
%School of Electrical and Computer Engineering,Network for
%Computational Nanotechnology,Purdue University, West Lafayette,
%Indiana,USA,47906 % \textbackslash\textbackslash
%}%
%\altaffiliation[Also at ]{Physics Department, XYZ University.}
\address{School of Electrical and Computer Engineering,Network for
Computational Nanotechnology,Purdue University, West Lafayette,
Indiana,USA,47906}
%\ead{custserv@iop.org}

\author{Timothy B.~Boykin }
%\affiliation{
%University of Alabama in Huntsville,Huntsville, Alabama 35899 USA % with \\
%}%
\address{University of Alabama in Huntsville,Huntsville, Alabama 35899 USA}

%\address{IOP
%Publishing, Dirac
%House, Temple Back, Bristol BS1 6BE, UK}
%\ead{custserv@iop.org}

\begin{abstract}
%https://en.wikipedia.org/wiki/Low-rank_approximation
The Empirical Tight Binding(ETB) method is widely used in atomistic
device simulations. The reliability of such simulations depends very
strongly on the choice of basis sets and the ETB parameters. The
traditional way of obtaining the ETB parameters is by fitting to
experiment data, or critical theoretical bandedges and symmetries
rather than a foundational mapping. A further shortcoming of
traditional ETB is the lack of an explicit basis.  In this work, a
DFT mapping process which constructs TB parameters and explicit
basis from DFT calculations is developed. The method is applied to
two materials: GaAs and MgO. Compared with the existing TB
parameters, the GaAs parameters by DFT mapping show better agreement
with the DFT results in bulk band structure calculations and lead to
different indirect valleys when applied to nanowire calculations.
The MgO TB parameters and TB basis functions are also obtained
through the DFT mapping process.
\end{abstract}
\maketitle
%\section{Introduction}
Modern semiconductor nanodevices have reached critical device
dimensions in the range of several nanometers. These devices consist
of complicated two and three dimensional geometries composed of
multiple materials. Typically, about 10000 to 10 million atoms are
in the active device region with contacts controlling the current
injection. This finite extent suggests an atomistic, local and
orbital-based electronic structure representation. Quantitative
device design requires the reliable prediction of bandgaps and band
offsets within a few meV and effective masses at principal symmetry
points within a few percent.

Ab-initio methods that have no adjustable parameters offer such
atomistic representations. However, accurate models such as hybrid functionals~\cite%
{HSE06}, GW~\cite{Hybertsen_GW} and BSE
approximations~\cite{BSE_PRL} are computationally far too expensive
to be applied on multi-million atom devices. More approximate
ab-initio methods such as the local density approximations (LDA) and generalized gradient approximations (GGA)~\cite%
{RMP_DFT} do not reproduce band gaps, relative band offsets, and
effective masses accurately enough. Empirical methods such as the
empirical tight binding (ETB) method are numerically much more
efficient. The accuracy of ETB is hereby limited by the parameters
fitting. Previous ETB simulations in semiconductor nanodevices such
as resonant tunneling diodes~\cite{Klimeck_RTD}, quantum
dots~\cite{Klimeck_QuantumDot} and strained Si/SiGe quantum
wells~\cite{Valleysplitting_Boykin} showed quantitative agreement
with experiments.

The accuracy of the ETB method depends critically on the careful
calibration of the empirical parameters. The common way to determine
the ETB parameters is to fit ETB results to experimental band structures~\cite%
{Jancu_Tightbinding}~\cite{Boykin_TB_strain}. %~(cite SL 1 and 2, J36, and
One shortcoming of this method is its requirement of experimental
data that are often not available for new and exotic materials. In
addition, the ETB basis functions remain unknown, which makes it
notoriously difficult to predict wave function dependent quantities
with high precission. To overcome these shortcomings, some
approaches were developed to construct localized
basis functions from ab-initio results such as localized wannier functions~%
\cite{Marzari_Wannier_functions} or quasi-atomic orbitals~\cite%
{Qian_quasiatomic_orbitals}~\cite{Lu_quasiatomic_orbitals}~\cite%
{Urban_TBfromDFT}. Unfortunately, these functions are either not
reliably centered at atoms, or resulting Hamiltonian requires long
distance coupling
with large number of neighbors, which is numerically expensive. %
Nanoelectronic devices are increasingly based on complex
heterstructures in 1 or 2 dimensions including atomistc disorder and
strain.  In this realm models with one or two neighbors are
conceptually preferrable and simple to implement.  In contrast the
validity of many neighbor bulk basis states are very questionable in
this domain.
% numerically expensive

In this work, a DFT mapping method that constructs ETB parameters
from ab-initio calculations is presented. This method allows to
determine ETB basis functions that are centered at atoms and it
limits the interatomic coupling to the first or second nearest
neighbors. Since the method does not require experimental results,
the ETB parameters are less empirical. Two materials are considered
in this work: 1) the well known GaAs to validate the method and 2)
MgO which is recently used as a magnetic tunneling barrier material
in Magnetoresistive Random-Access-Memory devices. MgO lacks the
elaborate experimental analysis, but it is known to have small
spin-orbit interaction and a large band
gap~\cite{Fe_MgO_Fe_junctions}.

%\section{Method}
\begin{figure}[h]
\center
\includegraphics[width = 9cm]{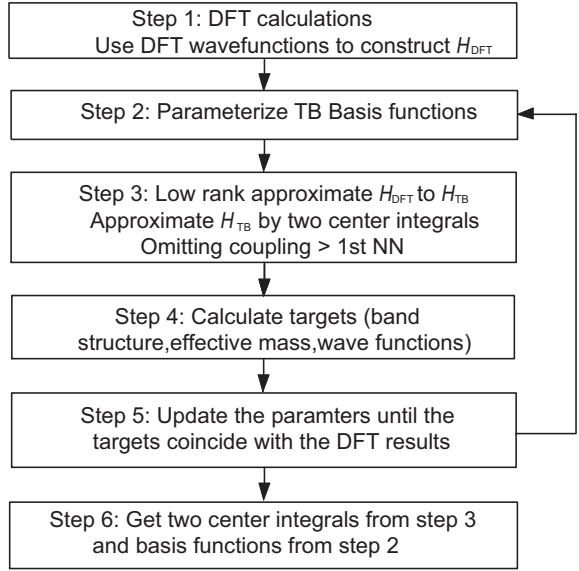} % Here is how to import EPS art% flowchart_v4
\caption{The process of TB parameters construction from DFT
calculations.} \label{fig:fig_flowchart}
\end{figure}

Figure~\ref{fig:fig_flowchart} shows the flow chart of the %density function theory (DFT)
mapping method. The first step is to perform ab-initio calculations
of the band structure of a material. In general, any method that is
capable to calculate electronic band structures and wave functions
is suitable here. However, in this work, DFT calculations including
hybrid functionals corrections for 60 electronic bands performed
with VASP version 5.2~\cite{Vasp_HSE} are used. In%~\cite{Vasp_GW}
the second step, the ETB basis functions for each type of atom are
defined as
\begin{equation}
\Psi _{n,l,m}\left( \mathbf{r}\right) \equiv \Psi _{n,l,m}\left(
r,\theta ,\phi \right) =R_{n,l}\left( r\right) Y_{l,m}\left( \theta
,\phi \right) , \label{eq:definition_atomic_orbitals}
\end{equation}%
where the functions $Y_{l,m}$ are the complex spherical harmonics
and the
functions $R_{n,l}$ are exponentially damped plane waves%
\begin{equation}
R_{n,l}\left( r\right) =\sum_{i=1}^{N}\left[ a_{i}\sin \left( {\lambda _{i}r}%
\right) {+b_{i}\cos }\left( {{\lambda _{i}r}}\right) \right]
r^{n-1}\exp \left( -\alpha _{i}r\right) .
\label{eq:definition_basis_Radialpart}
\end{equation}%
Here, $\mathbf{r=}\left( r,\theta ,\phi \right) $ is the position
vector with the respective atom centered at the origin, $l$, $m$ are
the angular and magnetic quantum numbers of the orbital basis
function, and $n$ is the principal number of the atomic orbital as
used in H\"{u}ckel type basis functions~\cite{Cerd_ETH}. The
remaining parameters $a_{i},b_{i},\alpha _{i},\lambda _{i}$ are the
fitting parameters. The ETB basis functions are spin independent.
With a given set of ETB basis functions $\Psi _{TB}^{\mathbf{k}}$,

%%%$\Psi _{\text{TB}}^{\mathbf{k}}$,

in the third step, a transformation matrix between these chosen DFT
basis functions %$\psi _{DFT}^{\mathbf{k}}$
$\psi_{\textrm{DFT}}^{\mathbf{k}}$ and the %$\Psi _{TB}^{\mathbf{k}}$
$\Psi _{ \textrm{TB}}^{\mathbf{k}}$ is calculated. Since the number
of the ETB basis functions is smaller than the DFT basis funcitons,
this transformation matrix is rectangular and represents a low rank
approximation.~\cite{LRA} Then, the DFT Hamiltonian is transformed
to the tight binding representation. The ETB Hamilton matrix
elements are approximated by two center integrals according to the
Slater-Koster
table~\cite{Slater_Tightbinding}~\cite{Podolskiy_TBElements}. Any
non-zero off-diagonal element of the overlap matrix is neglected.
ETB Hamilton matrix elements beyond either 1st or 2nd nearest
neighbor coupling are neglected. In Step 4, the band edges,
effective masses and eigen functions of the Hamiltonian at high
symmetry points are calculated and compared to the corresponding DFT
results. The overlaps of the ETB basis functions are also
determined. In the fifth step, all fitting parameters are adjusted
to improve 1) the agreement of the ETB\ results with the DFT ones
and 2) reduce the overlap matrix of the ETB basis functions to the
unity matrix. Steps 2 - 5 are repeated until the convergence
criterion is met, i.e. when the maximum difference of DFT and ETB band edges are within $%
10~\mathrm{meV}$, when the effective masses agree within 5\% and
when the eigenfunctions of DFT and ETB calculations agree by at
least 90\%. Step 6 requires to extract the converged ETB basis
functions and the ETB\ two center integrals.

\begin{figure}[h]
\center
\includegraphics[width = 11cm]{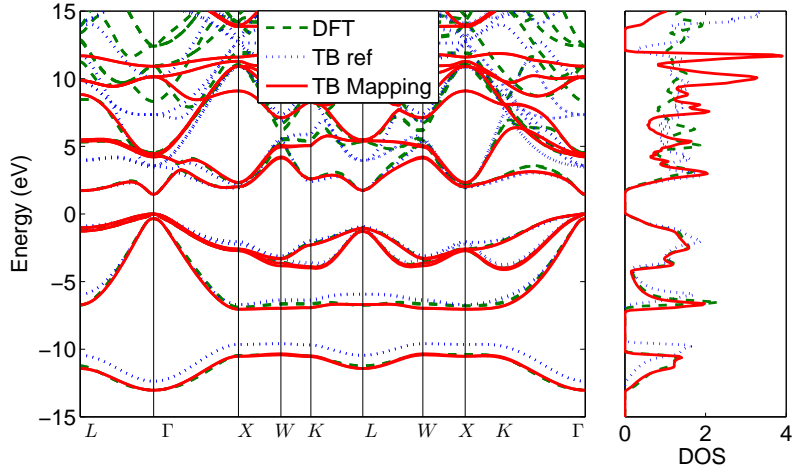}%
% Here is how to import EPS art
\caption{Band structure and density of states of GaAs by DFT(green
dashed lines), TB using parameters in
ref\protect\cite{Boykin_TB_strain} and TB using parameters by this
work.} \label{fig:fig_GaAsBand}
\end{figure}
For GaAs and MgO, eigenfunctions and eigen energies of the lowest 16
bands of the DFT calculations from $L$ to $\Gamma $ and $\Gamma $ to
$X$ are taken into account for the ETB fitting method. Here, the
wave functions for the topmost valence bands and lowest conduction
band valley are considered in the fitting of ETB eigenfunctions to
the DFT ones. The overlap of the ETB basis functions is partly
minimized in the fitting process. Most of the overlap matrix
elements vanish, but the maximum overlap i.e. in this case
the overlap of the $p$ orbitals of cations and $d$ orbitals of anions, e.g. $%
\langle p_{x}^{Ga}|d_{yz}^{As}\rangle =0.86$ at the $\Gamma $ point
remains comparably high.
\begin{figure}[t]
\center
\includegraphics[width = 10cm]{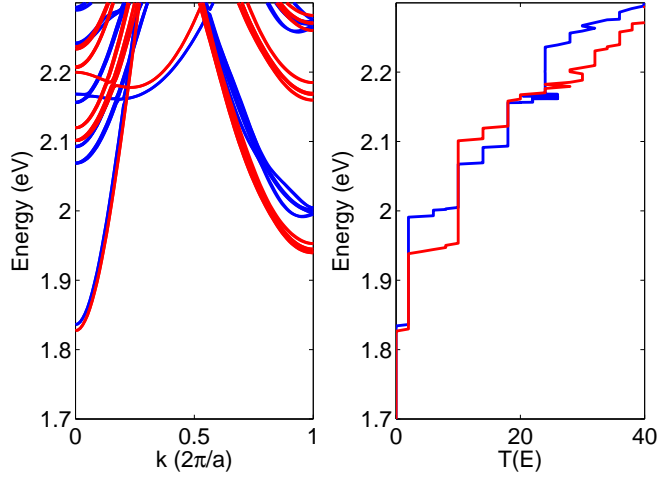}%
% Here is how to import EPS art
\caption{Comparison of E-k diagram of conduction bands (left figure)
and Transmission (right figure) of GaAs nanowire . The red lines are
results using TB parameters generated by this work; the blue line
are results using TB parameters by
ref\protect\cite{Boykin_TB_strain}.} \label{fig:fig_Ek_TE_GaAs}
\end{figure}
\begin{table}[t]
%\begin{ruledtabular}
\begin{tabular}{c|cccc|ccc}
\hline
&  &  & GaAs &  &  & MgO &  \\
Quantity & DFT & TB & Error & ref\cite{Boykin_TB_strain} & DFT & TB
& Error
\\ \hline
$E_g(\Gamma)$ & 1.420 & 1.449 & $2.96\%$ & 1.413 & 7.831 & 7.499 &
$4.2\%$
\\
$E_g(X)$ & 1.973 & 1.947 & $0.9\%$ & 1.898 & 12.161 & 11.819 & $2.8\%$ \\
$E_g(L)$ & 1.728 & 1.718 & $0.6\%$ & 1.714 & 10.871 & 10.469 & $3.7\%$ \\
$m^*_{\Gamma}$ & 0.0692 & 0.0737 & $6.5\%$ & 0.0657 & 0.396 & 0.458 & $%
15.6\% $ \\
$m^*_{(X,l)}$ & 1.140 & 1.117 & $2.0\%$ & 1.881 & $-$ & $-$ & $-$ \\
$m^*_{(X,t)}$ & 0.219 & 0.231 & $5.5\%$ & 0.175 & $-$ & $-$ & $-$ \\
$m^*_{(L,l)}$ & 1.700 & 1.756 & $3.3\%$ & 1.728 & $-$ & $-$ & $-$ \\
$m^*_{(L,t)}$ & 0.133 & 0.138 & $3.8\%$ & 0.098 & $-$ & $-$ & $-$ \\
$\gamma_1 $ & 6.964 & 6.985 & $1.1\%$ & 7.388 & 0.952 & 0.889 & $6.6\%$ \\
$\gamma_2 $ & 2.084 & 2.151 & $3.6\%$ & 2.367 & 0.277 & 0.219 & $20.9\%$ \\
$\gamma_3 $ & 2.972 & 2.980 & $1.1\%$ & 3.098 & 0.376 & 0.234 & $37.7\%$ \\
\hline
\end{tabular}%
\caption{Comparison of important bandedges and effective masses of
GaAs and MgO by DFT and TB.} \label{tab:Target_comparison}
\end{table}GaAs is parameterized for the 1st nearest neighbor $sp^3d^5s^*$ ETB
model. The resulting parameters can be found in table
\ref{tab:Parameter_GaAs}.
%Ref~\cite{ETB_Para}.
The band structure and density of states of GaAs are shown in Fig.~\ref%
{fig:fig_GaAsBand}. Calculated bandstructures of the DFT method, of
the ETB method with parameters of the present mapping method and of
the ETB method with previously published
parameters~\cite{Boykin_TB_strain} are compared in
Fig.~\ref{fig:fig_GaAsBand} as well. The ETB calculations with new
parameters agree very well with the DFT results for energies below
$5$~$\mathrm{eV}$ whereas ETB calculations with previously published
parameters deviate from the DFT results already at about
$2$~$\mathrm{eV}$. Relevant band edges and effective masses are
compared in
table~\ref{tab:Target_comparison}, demonstrating a much better fit. %Figure~\ref{fig:fig_GaAsBand} and
%table~\ref{tab:Target_comparison} show that ETB calculations using
%the present parameters agree better with DFT calculation than when
%the previously published parameters are used.

Figure~\ref{fig:fig_Ek_TE_GaAs} compares results of ETB calculations
of the band structure and the transmission coefficient of a GaAs $3.5~\mathrm{nm}%
\times 3.5~\mathrm{nm}$ squared nanowire when the new and the
previously published parameter set are used. The results agree for
the $\Gamma $ point at lower energies, but they deviate
significantly for the indirect conduction valley at about
$1.9~\mathrm{eV}$. The difference in the confinement energy of this
conduction valley originates from different transverse effective
masses at the L point of the two ETB parameter sets (see Table~\ref%
{tab:Target_comparison}). The modeling go the details of the high
symmetry points is particularly important for the emerging concept
of gamma-L transistors~\cite{LValley_Device}.
\begin{figure}[t]
\center
\includegraphics[width = 11cm]{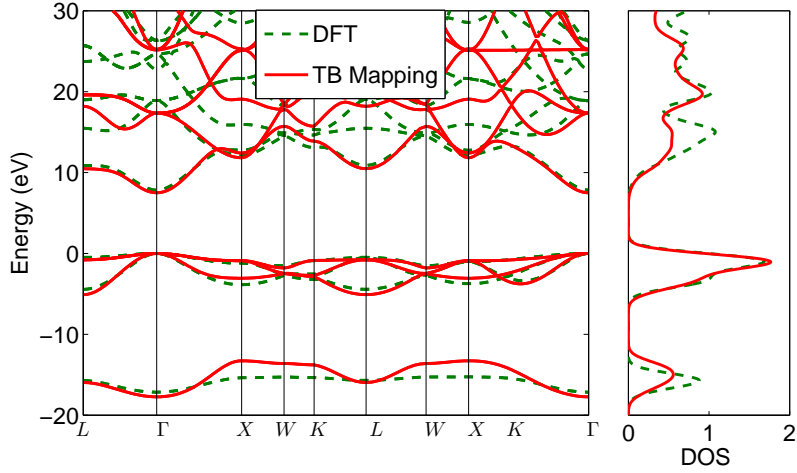}%
% Here is how to import EPS art
\caption{Band structure and density of states of MgO by DFT and TB.}
\label{fig:fig_Ek_TE_MgO}
\end{figure}

\begin{figure}[t]
\center
\includegraphics[width = 9cm]{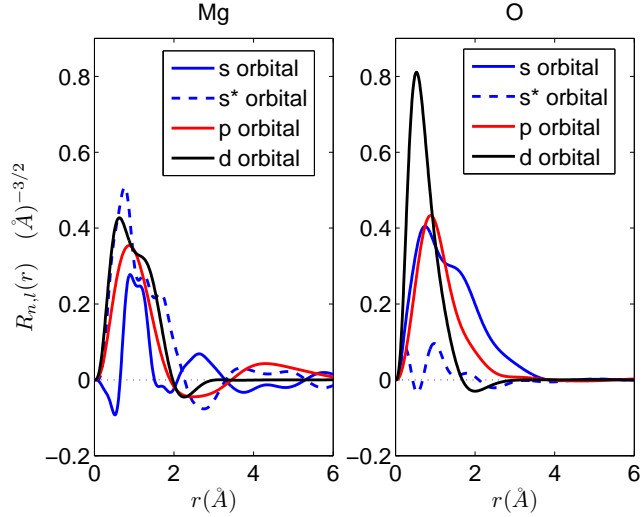}% Here is how to import EPS art
\caption{Radial part of Basis functions of Mg and O atoms in MgO.}
\label{fig:MgOBasisfunctions}
\end{figure}
MgO cyrstalizes in Rocksalt structure. Each Oxygen atom has 6
Magnesium atoms as 1st nearest neighbors and 12 Oxygen atoms as 2nd
nearest neighbors. The valence bands of MgO are formed by hybridized
orbitals of Oxygen atoms:
The $s$-orbitals and $p$-orbitals of Oxygen contribute to valence bands around $%
-17~\mathrm{eV}$ and $-1~\mathrm{eV}$ respectively, while the
$s^{\ast }$, $d$-orbitals of Oxygen and orbitals of Magnesium
contribute to the band structure for energies beyond $7~\mathrm{eV}$ (see Fig.~\ref%
{fig:fig_Ek_TE_MgO}). According to this, MgO is parameterized for a
2nd nearest neighbor $sp^3d^5s^*$ ETB\ model (for parameters see table \ref{tab:Parameter_MgO}). Within %~\cite{ETB_Para}
this model, the interaction between two Oxygen atoms is required to
produce the correct valence bands. The interaction between two
Magnesium atoms is omitted, since the onsite energies of Magnesium orbitals are higher than $%
20~\mathrm{eV}$ so that the omission will only affect bands close to
that high and technically irrelevant energy. The ETB band structure
matches the DFT result well within the energies $5$ to
$20~\mathrm{eV}$. Important band properties calculated in the DFT
and the ETB method are listed in Table~\ref{tab:Target_comparison}.
The basis functions of Oxygen and Magnesium are shown in Fig.~\ref%
{fig:MgOBasisfunctions}. The basis functions of Oxygen are more
localized while the basis functions of Magnesium are more plane wave
like functions. This difference in the localization originates from
the fact that Magnesium orbital energies are higher than the Oxygen
ones.

In conclusion, a method to determine ETB parameters from density
function theory calculations is developed. The method is applied to
GaAs and MgO. First nearest neighbor ETB parameters and basis
functions for an $sp^3d^5s^*$ model of GaAs are presented. Results
of this parameterization agree well with the DFT calculations. The
new ETB parameters lead to lower indirect conduction bands when
applied to GaAs nanowires. Second nearest neighbor ETB parameters
and basis functions for an $sp^3d^5s^*$ model of MgO are also
obtained. The ETB\ results with this parameterization also agree
well with the DFT calculations.

nanoHUB.org computational resources operated by the Network for
Computational Nanotechnology funded by NSF are utilized in this
work. The research was funded by the Lockheed Martin Corporation and
NSF (Award No. 1125017)

\begin{table}[h]
\caption{\label{tab:Parameter_GaAs} Tight Binding Parameters for
bulk GaAs}
%\begin{ruledtabular}
\begin{tabular}{|cc|cc|}
%Left\footnote{Note a.}&Centered\footnote{Note b.}&Right\\
\hline
  Parameter & value & Parameter & value  \\
\hline
$E_{sa}$ & $-4.5863$ & $s^{*}_ap_c\sigma$ & $2.6877 $\\
$E_{pa}$ & $1.4694$ & $s^{*}_cp_a\sigma$ & $1.8335 $ \\
$E_{s*a}$ & $10.0480$ & $s_ad_c\sigma$ & $-2.1172$ \\
$E_{da}$ & $11.2878$ & $s_cd_a\sigma$ & $-2.9128$ \\
$E_{sc}$ & $-1.3323$ & $s^{*}_ad_c\sigma$ & $-0.4974 $\\
$E_{pc}$ & $9.5885$ &  $s^{*}_cd_a\sigma$ & $-2.9971$\\
$E_{s*c}$ & $25.6752$ & $pp\sigma$ &  $3.8065 $\\
$E_{dc}$ & $35.2863$ & $pp\pi$ & $-1.5010$ \\
$\Delta_{a}$ & $0.1259 $  & $p_ad_c\sigma$ & $-1.2077$ \\
$\Delta_{c}$ & $0.1235$  & $p_cd_a\sigma$ & $-1.9855$ \\
$ss\sigma$ &  $-1.7615$ & $p_ad_c\pi$ & $3.1547$ \\
$s^{*}s^{*}\sigma$ &  $-0.8374$ & $p_cd_a\pi$ & $2.3234$ \\
$s_a^{*}s_c\sigma$ & $-1.1173$ & $dd\sigma$ & $-1.9986$ \\
$s_as_c^{*}\sigma$ & $-2.9313$ & $dd\pi$ & $3.1681$  \\
$s_ap_c\sigma$ & $2.1768 $ & $dd\delta$ & $-2.3137 $ \\
$s_cp_a\sigma$ & $3.6705 $ & & \\
\hline
\end{tabular}
%\end{ruledtabular}
\end{table}

 $$
 $$

\begin{table}[h]
\caption{\label{tab:Parameter_MgO} Tight Binding Parameters for bulk
MgO}
%\begin{ruledtabular}
\begin{tabular}{|cc|cc|}
%Left\footnote{Note a.}&Centered\footnote{Note b.}&Right\\
\hline
Parameter &  Value &  Parameter & Value \\
\hline
$E_{sa}$ & $-7.1496$ & $p_ad_c\pi$ & $1.5284$ \\
$E_{pa}$ &  $5.8926$ & $p_cd_a\pi$ & $-4.0453$ \\
$E_{s*a}$ & $23.8138$ & $d_ad_c\sigma$ & $-1.0038$ \\
$E_{da}$ & $40.0285$ & $d_ad_c\pi$ & $5.0830$ \\
$E_{sc}$ & $38.4754$ & $d_ad_c\delta$ & $-0.6323$ \\
$E_{pc}$ & $32.1465$ & $s_as_a\sigma$  &   $-0.2718$ \\
$E_{s*c}$ & $45.5084$ & $s^{*}_as_a^{*}\sigma$  &   $-0.4690$ \\
$E_{dc}$ & $57.9865$ & $s_as_a^{*}\sigma$  &   $-0.0001$ \\
$\Delta_{a}$ & $0.0031$  & $s_ap_a\sigma$ &    $0.3388$ \\
$\Delta_{c}$ & $0.0298$  & $s^{*}_ap_a\sigma$   & $0.1965$ \\
$s_as_c\sigma$ & $-0.1192$ & $s_ad_a\sigma$ & $-0.3380$ \\
$s^{*}_as^{*}_c\sigma$ &  $1.6477$ & $s^{*}_ad_a\sigma$   & $-0.4407$  \\
$s_a^{*}s_c\sigma$ & $-0.6008$ & $p_ap_a\sigma$   & $0.4371$ \\
$s_as_c^{*}\sigma$ & $-0.6347$ & $p_ap_a\pi$ &   $ -0.0641$ \\
$s_ap_c\sigma$ & $-2.0013$ & $p_ad_a\sigma$ &   $ -0.4039$ \\
$s_cp_a\sigma$ &  $0.3283$ & $p_ad_a\pi$ &   $0.6986$ \\
$s^{*}_ap_c\sigma$ & $2.1584$ & $d_ad_a\sigma$ &   $-2.3768$ \\
$s^{*}_cp_a\sigma$ & $2.1282$ & $d_ad_a\pi$    &   $0.4556$ \\
$s_ad_c\sigma$ &  $0.6641$ & $d_ad_a\delta$ &   $0.0967$ \\
$s_cd_a\sigma$ & $-2.9483$ & & \\
$s^{*}_ad_c\sigma$ & $1.6890$ &  &\\
$s^{*}_cd_a\sigma$ & $3.1534$ &  &\\
$p_ap_c\sigma$ &  $0.1743$ & & \\
$p_ap_c\pi$ & $-0.4703$ &  &\\
$p_ad_c\sigma$ & $-1.9960$ &  &\\
$p_cd_a\sigma$ & $0.0519$ &  & \\
\hline
\end{tabular}
%\end{ruledtabular}
\end{table}

\section*{References}
%\begin{thebibliography}{10}
%\bibitem{book1} Goosens M, Rahtz S and Mittelbach F 1997 {\it The \LaTeX\ Graphics Companion\/}
%(Reading, MA: Addison-Wesley)
%\bibitem{eps} Reckdahl K 1997 {\it Using Imported Graphics in \LaTeX\ } (search CTAN for the file `epslatex.pdf')
%\end{thebibliography}
%

\end{document}